\documentclass[pra,aps,byrevtex,showpacs,superscriptaddress,nofootinbib,twocolumn,letterpaper]{revtex4-1}
\usepackage{graphicx}
\usepackage{amsmath}
\usepackage{amssymb}
\usepackage{mathrsfs}
\usepackage{hyperref}
\usepackage{microtype}
\hypersetup{colorlinks=true,linkcolor=blue,citecolor=blue,urlcolor=blue}

\begin{document}

\title{Experimental Multi-Party Sequential State Discrimination}

\author{M.~A.~Sol\'is-Prosser}
\author{P.~Gonz\'alez}
\author{J.~Fuenzalida}
\author{S.~G\'omez}
\affiliation{Center for Optics and Photonics, Universidad de Concepci\'on, Casilla 4016, Concepci\'on, Chile}
\affiliation{MSI-Nucleus on Advanced Optics, Universidad de Concepci\'on, Casilla 160-C, Concepci\'on, Chile}
\affiliation{Departamento de F\'isica, Universidad de Concepci\'on, Casilla 160-C, Concepci\'on, Chile}

\author{G.~B.~Xavier}
\affiliation{Center for Optics and Photonics, Universidad de Concepci\'on, Casilla 4016, Concepci\'on, Chile}
\affiliation{MSI-Nucleus on Advanced Optics, Universidad de Concepci\'on, Casilla 160-C, Concepci\'on, Chile}
\affiliation{Departamento de Ingenier\'ia El\'ectrica, Universidad de Concepci\'on, 160-C, Concepci\'on 4070386, Chile}

\author{A.~Delgado}
\author{G.~Lima}
\affiliation{Center for Optics and Photonics, Universidad de Concepci\'on, Casilla 4016, Concepci\'on, Chile}
\affiliation{MSI-Nucleus on Advanced Optics, Universidad de Concepci\'on, Casilla 160-C, Concepci\'on, Chile}
\affiliation{Departamento de F\'isica, Universidad de Concepci\'on, Casilla 160-C, Concepci\'on, Chile}

\begin{abstract}
Recently, a protocol for quantum state discrimination (QSD) in a multi-party scenario has been introduced [\prl~{\bf 111}, 100501 (2013)]. In this protocol, Alice generates a quantum system in one of two pre-defined non-orthogonal qubit states, and the goal is to send the generated state information to different parties without classical communication exchanged between them during the protocol's session. The interesting feature is that, by resorting to sequential generalized measurements onto this single system, there is a non-vanishing probability that all observers identify the state prepared by Alice. Here, we present the experimental implementation of this protocol based on polarization single-photon states. Our scheme works over an optical network, and since QSD lies in the core of many protocols, it represents a step towards experimental multi-party quantum information processing.
\end{abstract}

\date{\today}
\maketitle

\section{Introduction}
The unavoidable change of quantum states by a measurement process is one of the main distinguishing features of quantum mechanics \cite{VonNeumann,Braginsky}. The acquisition of information about a quantum system through a projective measurement creates a post-measurement state which, when inspected, does not allow one to deduce the state of the system before the measurement. Thus, this class of measurement has a destructive character since all information about the initial state is lost. However, not all measurement processes lead to a complete loss of information on the initial state \cite{Scully,Bertet,Durr,Neves}.

The trade-off between the information about a quantum system, obtained by means of a measurement process, and the disturbance induced by this process in the state of the system has been an intensive research subject \cite{Banaszek,Englert,Sciarrino}. In this context, sequential measurements have been recently considered. In such case, several measurement processes are carried out consecutively onto the same physical system \cite{Filip}. Adopting sequential measurements, the existence of an adaptive strategy which optimises the balance between the acquired information and the disturbance induced in the state of the system has been experimentally demonstrated \cite{Nagali}.

In this work we present motivating fundamental results related with sequential measurements onto a single quantum system. We experimentally demonstrate that multi-party sequential unambiguous state discrimination (SUSD) is possible \cite{Bergou1}. The discrimination among nonorthogonal quantum states was first introduced in the context of quantum decision theory \cite{HelstromBook,Holevo73,Yuen75}. The idea is that a party, Alice, prepares a quantum system in one of several pre-defined nonorthogonal states. Then, she sends the system to a second party, Bob, whose task is to determine what is the state of the system. Several discrimination strategies are possible depending on the constraints imposed. Unambiguous state discrimination (USD) is designed upon the requirement of perfect identification of the nonorthogonal states considered \cite{IDP}. Due to the nonorthogonality, this requirement can only be achieved probabilistically. Thus, the scheme admits the possibility of an inconclusive event that does not give any information about Alice's prepared state.

Sequential unambiguous state discrimination arises in a scenario where the states prepared by Alice must be unambiguously identified by multiple parties, and without the use of classical communication exchanged between them during the protocol's session. Since the process must be carried out without sharing a classical bit of information, each observer has access only to the post-measurement unknown states generated by the previous one. Through a sequence of consecutive generalized measurements on the same quantum system, there is a non-vanishing probability that all receivers simultaneously identify the state prepared by Alice. The process of SUSD is designed to maximize this probability, and is described as the concatenation of non-optimal USDs carried out by the intermediate parties, to an optimal USD implemented by the last party.

\begin{figure*}[t]
\centering	
\includegraphics[width=0.7\textwidth]{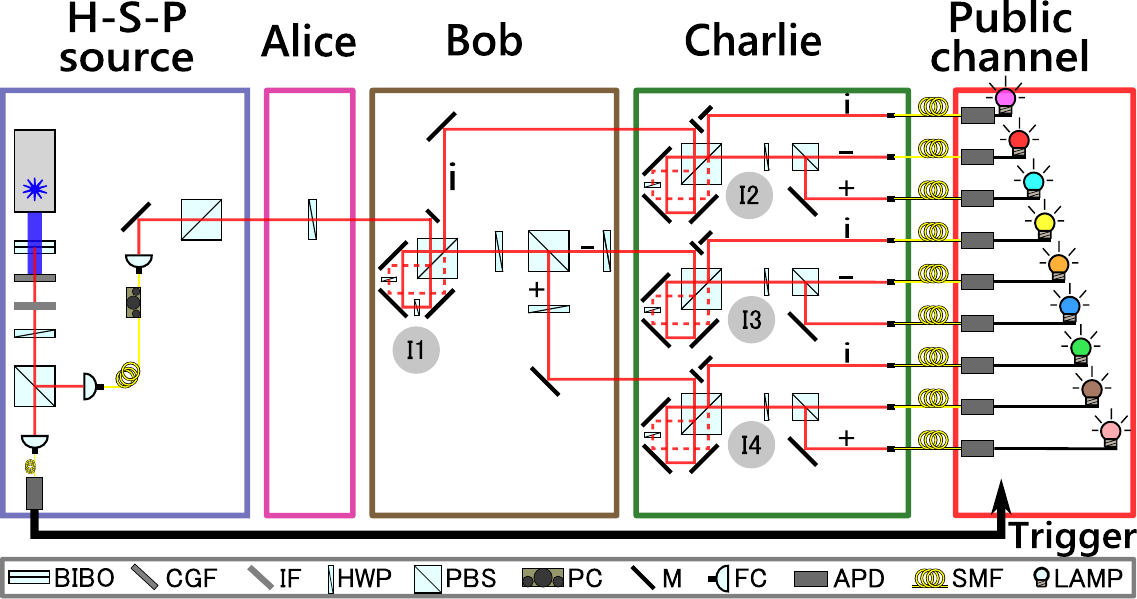}\\
\caption{(Color online) Experimental setup. Alice encodes the state $|\psi_{+}\rangle$ or $|\psi_{-}\rangle$, with a half-wave plate, on a down-converted photon heralded by the other twin comprising the heralded single-photon (H-S-P) source. Bob resorts to a Sagnac-like interferometer to implement the non-optimal USD measurement. Bob's post-measurement state is sent to Charlie, who then carries out an optimal USD procedure with three identical Sagnac-like interferometers. Each output i represents an inconclusive event, while the other two, + and -, are univocally associated to the states $|\psi_{+}\rangle$ and $|\psi_{-}\rangle$, respectively. The detectors are placed in a public channel, such that both Bob and Charlie can see which detector clicked in each run of the experiment. Therefore, the result of their individual measurement is obtained without the need of classical communication between the parties. CGF: colored glass filter, IF: interference filter (bandpass), HWP: half-wave plate, PBS: polarizing beam splitter, PC: polarization control module, M: mirror, FC: optical fiber coupler, APD: avalanche photo-detector, SMF: single mode fiber.  \label{fig:setup}}
\end{figure*}

Throughout the development of quantum information theory, the studies of protocols concerning discrimination of non-orthogonal quantum states have shown their close relationship with others such as quantum key distribution (QKD) \cite{Bennett92}, entanglement concentration \cite{Chefles97,Chefles98-1,Croke08,Yang09}, quantum cloning \cite{Duan98,Jimenez10}, teleportation \cite{Roa03,Neves12,SolisProsser13a}, entanglement swapping \cite{Delgado05,SolisProsser14}, superdense coding \cite{Pati05}, and some quantum algorithms \cite{Bergou03}. Thus, our experimental implementation of a SUSD can be seen as a step towards single-photon multi-party experimental quantum information processing. Moreover, our scheme is based on single-photon polarization qubit states and can be extended to have the parties separated over an optical network \cite{Guix1, Guix2, Xu}.

\begin{figure*}[t]
	\centering
	\includegraphics[width=0.8\textwidth]{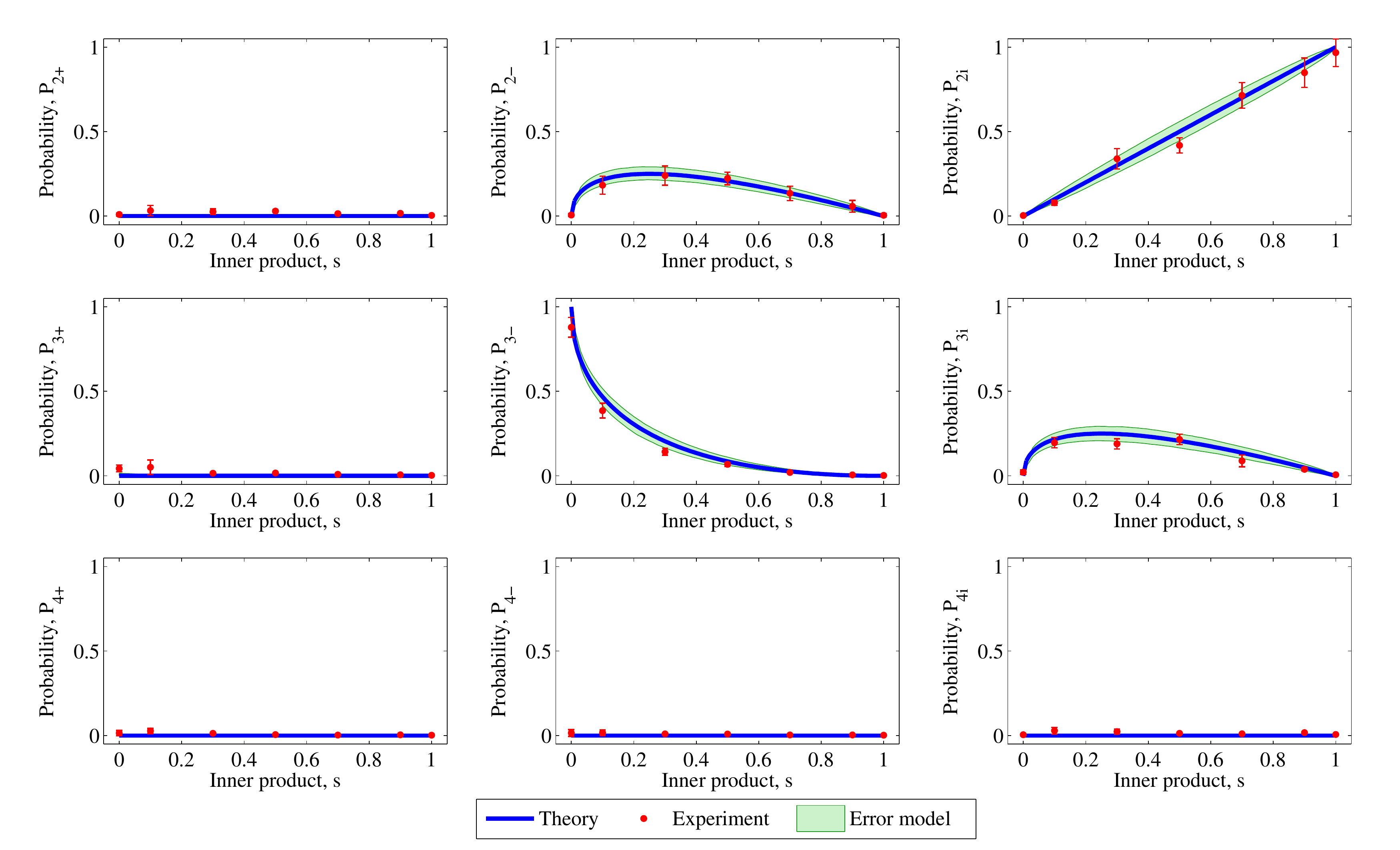}	
	\caption{(Color online) Probabilities $P_{\mu k}$ for each detection configuration while sending $|\psi_{-}\rangle$. The continuous line represents the theoretically expected values for each probability as a function of the inner product $s$. The points with error bars represent the recorded probabilities. The green area represents an error model (see text for details). \label{fig:psiminus}}
\end{figure*}

\section{Theory}

Now, let us briefly review the SUSD process between three parties (named Alice, Bob and Charlie) \cite{Bergou1}. Alice prepares a quantum system in a state randomly chosen from a set of two equally likely nonorthogonal states defined by $|\psi_\pm \rangle = a|h\rangle \pm b|v\rangle$. States $|h\rangle$ and $|v\rangle$ form a two-dimensional orthonormal basis and coefficients $a$ and $b$ are real and positive numbers such that $a^2+b^2=1$. Without loss of generality, we assume that $a\geqslant b$, so the inner product $s=\langle\psi_{+}|\psi_{-}\rangle =a^2-b^2$ is real and non-negative. Thereby, coefficients $a$ and $b$ can be considered as functions of $s$, that is $a^2 = \frac{1+s}{2}$, $b^2 = \frac{1-s}{2}$. After state preparation Alice sends the quantum system acting as an information carrier to Bob. He now implements a non-optimal USD process. This allows Bob to obtain information about the state prepared by Alice, while leaving enough information in the post-measurement states such that Charlie can also discriminate Alice's states. This happens at the expense of a reduced success probability. The non-optimal USD performed by Bob requires the implementation of a specific generalized quantum measurement. According to the Neumark's theorem, such measurement can be implemented by resorting to an additional ancillary system, an entangling unitary operation between the ancilla and the system, and finally by a projective measurement over the ancilla Hilbert space (please see the \hyperref[sec:app]{Appendix} for more details). In terms of a bipartite unitary operation between the system and the ancilla, Bob's discrimination process is described as
\begin{align}
|\psi_\pm\rangle|B\rangle \longrightarrow \sqrt{1-\sqrt{s}}|\pm\rangle|b_0\rangle ~\mp ~ s^{1/4}|\phi_\pm \rangle|b_1\rangle,\label{eq:Bob}
\end{align}
where $|B\rangle$ is an arbitrary state representing the initial state of the ancilla and $\{|b_0\rangle ,|b_1\rangle\}$ is a two-dimensional orthonormal basis of the ancilla. The states $|\pm\rangle$ and $|\phi_\pm\rangle$ are given by $|\pm\rangle = \frac{1}{\sqrt{2}}[|h\rangle\pm|v\rangle]$ and $|\phi_\pm\rangle = \sqrt{\tfrac{1}{2}(1-\sqrt{s})}|h\rangle \mp \sqrt{\tfrac{1}{2}(1+\sqrt{s})}|v\rangle$, respectively. Thus, Bob implements a process that transforms the set of nonorthogonal states into orthogonal ones, succeeding with probability $1-\sqrt{s}$. Since the states are now orthogonal, Bob can discriminate between them and deduce the state prepared by Alice. Otherwise, there exists a probability $\sqrt{s}$ of obtaining no information. After the implementation of Bob's USD, the particle is sent to Charlie who does not know the result of Bob's discrimination. In consequence, the state of the particle cannot depend on the particular outcomes obtained by Bob and must be always $|\phi_\pm\rangle$ regardless of his recorded results. Charlie discriminates unambiguously between these two states whose inner product is now $-\sqrt{s}$, and therefore, less separated than the original states $|\psi_\pm\rangle$ because $|\langle \phi_{-}|\phi_{+}\rangle|\geqslant \langle \psi_{-}|\psi_{+}\rangle$. Analogously, Charlie must set up an operation such that
\begin{align}
	|\phi_\pm\rangle|C\rangle \longrightarrow \sqrt{1- \sqrt{s}}|\pm \rangle|c_0\rangle \mp s^{1/4}|h\rangle|c_1\rangle ,\label{eq:Charlie}
\end{align}
where $|C\rangle$ is the initial state of the ancilla used by Charlie. $\{|c_0\rangle,|c_1\rangle\}$ corresponds to its possible final orthogonal states that inform Charlie whether the discrimination has been successfully achieved or not. Charlie's USD must be optimal since no additional parties will handle the system and, therefore, the remaining information must be maximally acquired. The state of the system in case of an inconclusive event is the same one despite the state Alice sent \cite{IDP}. As seen from Eq.~(\ref{eq:Charlie}), Charlie has a ${1-\sqrt{s}}$ probability for succeeding and $\sqrt{s}$ for failing. Taking into account Eq.~(\ref{eq:Bob}), we have that
\begin{align}
	P_{\rm succ} = (1-\sqrt{s})^2,\label{eq:psucc}
\end{align}
is the probability of both Bob and Charlie having successfully accomplished their quantum state discrimination, which is the main purpose of this protocol. Note that Eq. (\ref{eq:psucc}) is the upper limit for the joint success probability of the SUSD protocol \cite{Bergou1}.

\section{Experiment}

In our implementation, the states $|\psi_\pm\rangle$ are encoded in the polarization degree of freedom of a single photon. Figure~\ref{fig:setup} illustrates our experimental setup. A 60 mW CW laser at 355 nm is used to generate twin photons through spontaneous parametric down conversion in a non-linear ${\rm Bi B_3 O_6 }$ (BiBO) crystal. A colored glass filter and an interference filter (10 nm bandwidth) block the pump field and select degenerated down-converted photons at 710 nm. A heralded single-photon source uses the detection of a down-converted photon to witness the passage of the other twin photon through the experimental scheme. The latter is sent through a single mode optical fiber (SMF) to eliminate any spatial correlation between the twin photons, and then through a polarizing beam splitter (PBS) for polarization filtering. This source produces a typical coincident count rate of {$\sim 2600/$s} with an accidental count rate of $\sim 15$/s. The counts are recorded by a Field Programmable Gate Array coincidence counting unit, with the timing delay adjusted between each detector's output and the heralding trigger signal.

Alice prepares one of the states $|\psi_\pm\rangle$ with a single half-wave plate (HWP). Bob's non-optimal USD is implemented by means of an intrinsically stable Sagnac interferometer I1 (see Fig. \ref{fig:setup}). The propagation path of a photon within I1 depends on its polarization state, which allows for conditional polarization transformations implemented with half-wave plates placed inside the interferometer \cite{Nagali,Fabian}. The fast axes of the plates located at the clockwise transmitted mode (continuous line in I1), and at the counter-clockwise reflected path (dashed line in I1), were oriented at angles $\frac{1}{2}\arccos\sqrt{ \frac{1-\sqrt{s}}{1+s} }$ and $\frac{1}{2}\arccos\sqrt{ \frac{1}{1+\sqrt{s}} } +\frac{\pi}{2}$, respectively. The upper and lower output ports of the interferometer I1 are associated to a failure or a success of Bob's discrimination process, respectively. In such configuration, a photon emerging from the upper output port is described by one of the two nonorthogonal states $|\phi_{\pm}\rangle$. A photon emerging from the lower output port is described by one of two orthogonal polarization states $|\pm\rangle$. These states can be deterministically discriminated by splitting them into two new paths using a PBS and a HWP. Since they are univocally related to states $|\psi_\pm\rangle$, this identifies with certainty the states prepared by Alice. This concludes the generalized measurement of Bob implementing the USD process.

Before the transmission of the photon to Charlie's I3 and I4 interferometers, Bob transforms the post-measurement state $|+\rangle$ to $|\phi_{+}\rangle$, and $|-\rangle$ to $|\phi_{-}\rangle$ using HWPs (see Fig. \ref{fig:setup}). Thus, the state of the photon sent to Charlie is always one of the polarization states $|\phi_{\pm}\rangle$, independently of the path followed after Bob's discrimination (in this way, there is no bit of classical communication being shared by them). Charlie analyses the polarization state by placing new Sagnac interferometers I2, I3 and I4 at the end of each path leaving Bob's measurement. These three interferometers are similar to Bob's one and their settings are chosen to achieve the optimal USD of $|\phi_{\pm}\rangle$ in Charlie's measurement process \cite{Fabian}. For the optimal USD, Charlie needs to place a HWP in each counter-clockwise mode in the interferometers I2, I3, and I4 of Fig.~\ref{fig:setup}. They must be adjusted at an angle given by: $\frac{1}{2}\arccos\sqrt{\frac{1-\sqrt{s}}{1+\sqrt{s}}}$. Another HWP oriented at $0$ degrees was used in the clockwise mode to balance the interferometers. Note that Charlie must adopt three interferometers, as he does not know which path corresponds to a conclusive or inconclusive result of Bob's measurement. Last, it is worth mentioning that such setup configuration can find application to multipartite secure quantum communication since, for each trial, Bob can guarantee that there is no flow of information - regarding his measurement - to Charlie by randomizing at which output path the post-measurement polarization states are sent in. The same is valid for Charlie, who can guarantee the privacy of his measurements by randomizing for each interferometer, the detectors corresponding to the conclusive and inconclusive events.

Finally, the photon is measured with silicon avalanche photo-detectors (APDs) operating in continuous mode, with a detection efficiency of $\sim 60\%$ coupled with single-mode fibers (SMFs). For each detection event, the clicking detector is publicly announced. From the detection events, one can estimate when Bob, Charlie or both succeed discriminating the state generated by Alice. For instance, consider that the last detector was the one that clicked. Based on this publicly announced information, both Bob and Charlie get informed that the state prepared by Alice was $|\psi_\pm\rangle$, without the need of using any classical communication exchanged between the three parties. Last, please note that in our implementation Bob and Charlie don't need to share one bit of classical information to define how Bob's measurement outcomes map to the outcomes of Charlie's measurement (or, alternatively, to the detectors at the public channel). There is a simple strategy that Bob can adopt to figure out the mapping. That is, Bob can simply block two of his outputs and note from the detection events publicly announced, which detectors are associated with the unblocked path. The same strategy is also valid for Charlie and the public detectors.
\begin{figure}[t]
	\centering
	\includegraphics[width=0.95\columnwidth]{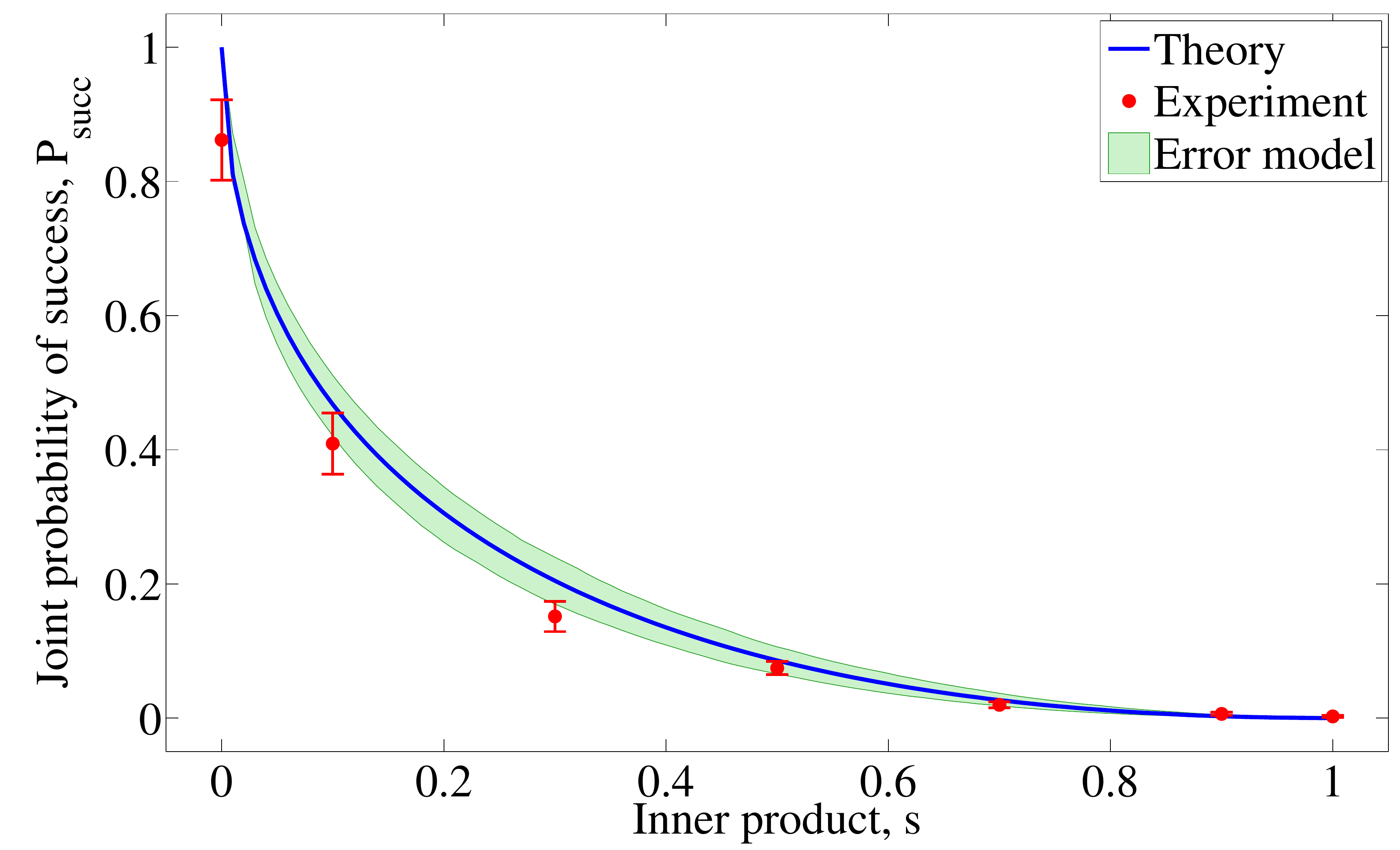}\\
	\caption{(Color online). Joint success probability of Bob and Charlie succeeding at their discrimination attempts as a function of $s$. The continuous line represents the theoretical prediction of Eq.~(\ref{eq:psucc}). The points with error bars represent the recorded probabilities averaged over $|\psi_{+}\rangle$ and $|\psi_{-}\rangle$. The green area represents an error model to show how experimental imperfections can affect $P_{succ}$ in our scheme.
  \label{fig:total_success}}
\end{figure}

To experimentally validate our scheme, the SUSD protocol was implemented considering seven different values for the inner product ($s$) of the non-orthogonal states generated by Alice. The results when Alice sends $|\psi_{-}\rangle$ are shown in Fig.~\ref{fig:psiminus}, as an illustrative example of the results recorded. Each plot shows the detection probability $P_{\mu k}$, with $ \mu=2,3,4$ and $k=+,-,i$, at the output of the $\mu$th interferometer (see Fig.~\ref{fig:setup}). These probabilities $P_{\mu k}$ are the observed average values after 45 experimental runs, each consisting of 15 s integration time, for each value of $s$. The error bars display the standard deviation of each probability considering all experimental runs. The continuous line represents the theoretically predicted values for each probability as a function of the inner product $s$. A good agreement between the theory and experimental results can be observed. To understand how experimental imperfections can affect our proposed scheme, we resorted to an error model based on Monte Carlo simulations, which is represented by a green area. To account for errors in Alice's state preparation procedure we used a measured maximum mismatch of $\pm 1^{\circ}$ on the half-wave plates. Furthermore we also considered in the model the use of non-ideal PBSs, where up to 3\% of horizontally (vertically) polarized light could be lost while being transmitted (reflected) through the PBS, and a mismatch between the output modes of the Sagnac-like interferometers of up to 3\%. Fig.~\ref{fig:total_success} shows the recorded joint success probability (averaged over $|\psi_{+}\rangle$ and $|\psi_{-}\rangle$) for each value of $s$. Again, there is a good agreement between theory [Eq.~(\ref{eq:psucc})] and the experimental results observed.

\section{Conclusions}
Here we reported on the implementation of the recently proposed protocol of sequential unambiguous state discrimination aimed at quantum information processing in a multi-party scenario \cite{Bergou1}. In our scheme, a single photon is used to distribute the information of a qubit polarization state between three parties: Alice, Bob and Charlie, all connected through a linear optical network. The scheme relies on the generation of non-orthogonal polarization states by Alice, with Bob and Charlie applying sequential measurements onto the single photon for the unambiguous state discrimination. Our experimental demonstration employs commercially available optical components, opening up a path towards practical applications. Finally, the technology to compensate polarization drifts in long-optical fibers is available \cite{Guix1, Guix2, Xu}, which opens up the possibility to spatially separate the parties with long-distance optical fiber links.

\acknowledgments
The authors thank Fabio~Sciarrino for useful discussions. This work was supported by the Chilean Grants FONDECYT~1160400, FONDECYT~1140635, Ministerio de Econom\'ia, Fomento y Turismo (ICM RC130001), and PIA-CONICYT~PFB0824. M.A.S.P. and P.G. acknowledge CONICYT for financial support.

\appendix*
\section{Remarks on the implementation of unambiguous state discrimination \label{sec:app}}

The unambiguous state discrimination (USD) protocol requires the implementation of a generalized quantum measurement over a given physical system and, in accordance with the Neumark's theorem, its implementation consists of three basic steps: (i) the addition of an ancillary system, (ii) a collective or joint unitary transformation between the system and the ancilla, and (iii) a projective measurement over the ancilla Hilbert space.

To see this, please let us consider that the nonorthogonal states to be discriminated $|\psi_\pm\rangle_s=a|0\rangle_s\pm b|1\rangle_s$ are encoded in a system $s$. With the purpose of performing the USD of such states, we must then consider that an extra ancillary system $b$, initially prepared in an arbitrary state $|B\rangle_b$, is added to this system. Last, let us assume that the following unitary transformation $U_{sb}$ acting onto system $s+b$ is implemented:
\begin{equation}
U_{sb}|\psi_\pm\rangle_s|B\rangle_b=\sqrt{p}|\pm\rangle_s|b_0\rangle_b+\sqrt{1-p}|\phi_{\pm}\rangle_s|b_1\rangle_b.
\label{Theo}
\end{equation}
Then, one can see that this transformation always creates a set of three mutually orthogonal states: the two states $|\pm\rangle_s|b_0\rangle_b=(1/\sqrt{2})(|0\rangle_s\pm|1\rangle_s)|b_0\rangle_b$ and one of the states $|\phi_{\pm}\rangle_s|b_1\rangle_b$. The states $|+\rangle_s|b_0\rangle_b$ and $|-\rangle_s|b_0\rangle_b$ are associated with the successful identification of the nonorthogonal states $|\psi_+\rangle_s$ and $|\psi_-\rangle_s$, respectively. The states $|\phi_{\pm}\rangle_s|b_1\rangle_b$ are associated with a failure in the identification process. Since the states $|+\rangle_s|b_0\rangle_b$, $|-\rangle_s|b_0\rangle_b$ and $|\phi_{\pm}\rangle_s|b_1\rangle_b$ are mutually orthogonal, they can be perfectly distinguished. Thus, conclusively identifying the two nonorthogonal states.

The three elements of the generalized measurement representing the USD process are, therefore, ${\Pi_m = A_{m}^{\dagger}A_{m}^{\,}}$, where $A_{m}$ are the Kraus operators describing the physical processes performed on the initial state of system $s$. These operators are given by
\begin{align}
    {}_s\langle k|A_m|j\rangle_s = {}_s\langle k|{}_b\langle b_m|U_{sb}|j\rangle_s|B\rangle_b,
\end{align} 
where $\{|j\rangle_s\}$ is the computational basis for system $s$ and $\{|b_m\rangle_b\}$ is the basis on which the ancillary system $b$ is measured.

Typical optical implementations of unambiguous state discrimination (or alternatively, optical implementations of generalized quantum measurements), such as the ones adopted by Bob and Charlie in our work, can be fully accomplished by resorting to different degrees of freedom of a single photon \cite{Huttner96,Clarke01,Fabian}. To understand it, please consider that the states $|0\rangle_s$ and $|1\rangle_s$ are now the horizontal and vertical polarization states $|h\rangle_s$ and $|v\rangle_s$ of a single photon, respectively. Correspondingly, the diagonal polarization states are defined by $|\pm\rangle_s=(1/\sqrt{2})(|h\rangle_s\pm|v\rangle_s)$.

Without loss of generality, let us now focus on Bob's measurement. Note, however, that the same reasoning is valid for Charlie's measurement. Immediately after the unitary transformation $U_{sb}$ implemented by Bob's interferometer, the state of a single photon is given by
\begin{equation}
U_{sb}(a|h\rangle_s\pm b|v\rangle_s)|B\rangle_b=\sqrt{p}|\pm\rangle_s|b_0\rangle_b+\sqrt{1-p}|\phi_{\pm}\rangle_s|b_1\rangle_b,
\label{Exp}
\end{equation}
where, here, the state $|B\rangle_b$ represents the propagation path of photons at the input port of Bob's interferometer. The state $|b_0\rangle_b$ represents the photons emerging from the lower output port of Bob's interferometer (See IC1 in Fig.~1), while state $|b_1\rangle_b$ describes photons emerging from the upper output port of Bob's interferometer. Thereby, the ancillary degree of freedom employed in our experiment corresponds to the propagation path of the photons. The identification of the states $|\psi_\pm\rangle_s$ is then obtained by mapping the polarization states $|\pm\rangle_s$ of the photons (at the lower output port) to distinct propagation path modes. This is achieved with the help of a half-wave plate and a polarizing beam splitter. The combined action of a half-wave plate and a polarizing beam splitter transforms the states $|\pm\rangle_s|b_0\rangle_b$ onto states $|h\rangle_s|b_2\rangle_b$ and $|v\rangle_s|b_3\rangle_b$, respectively. These can be perfectly discriminated by placing photo-detectors on paths $|b_2\rangle_b$ and $|b_3\rangle_b$.

Summarizing: the action of Bob's interferometer, followed by a half-wave plate and a polarizing beam splitter at lower output port, is to transform the states $|\psi_\pm\rangle_s=a|h\rangle_s\pm b|v\rangle_s$ into three mutually orthogonal states: $|\phi_{\pm}\rangle_s|b_1\rangle_b$, $|h\rangle_s|b_2\rangle_b$ and $|v\rangle_s|b_3\rangle_b$. In turn, this allows for the successful identification of states $|\psi_\pm\rangle_s$ by means of placing photo-detectors at paths $|b_1\rangle_b$, $|b_2\rangle_b$ and $|b_3\rangle_b$. Therefore, implementing the desired USD operation.

\end{document}